\documentclass[journal=jacsat,manuscript=article]{achemso}

\usepackage[version=3]{mhchem} 
\usepackage{xcolor}
\usepackage{hyperref}
\usepackage{cleveref}
\usepackage{amsmath}
\usepackage{amssymb}

\author{Kirill Shmilovich}
\affiliation[University of Chicago]
{Pritzker School of Molecular Engineering, University of Chicago, Chicago, Illinois 60637, United States}
\altaffiliation{Equal Contribution}
\author{Benson Chen}
\affiliation[Insitro]
{Insitro, South San Francisco, California 94080, United States}
\altaffiliation{Equal Contribution}
\author{Theofanis Karaletsos}
\affiliation[Insitro]
{Insitro, South San Francisco, California 94080, United States}
\email{theofanis@insitro.com}
\author{Mohammad M. Sultan}
\affiliation[Insitro]
{Insitro, South San Francisco, California 94080, United States}
\email{msultan@insitro.com}

\title{DEL-Dock: Molecular Docking-Enabled Modeling of DNA-Encoded Libraries}

\begin{document}

%
%
%
%
%

\newpage
\begin{abstract}
DNA-Encoded Library (DEL) technology has enabled significant advances in hit identification by enabling efficient testing of combinatorially-generated molecular libraries. DEL screens measure protein binding affinity though sequencing reads of molecules tagged with unique DNA-barcodes that survive a series of selection experiments. Computational models have been deployed to learn the latent binding affinities that are correlated to the sequenced count data; however, this correlation is often obfuscated by various sources of noise introduced in its complicated data-generation process. In order to denoise DEL count data and screen for molecules with good binding affinity, computational models require the correct assumptions in their modeling structure to capture the correct signals underlying the data. Recent advances in DEL models have focused on probabilistic formulations of count data, but existing approaches have thus far been limited to only utilizing 2-D molecule-level representations. We introduce a new paradigm, DEL-Dock, that combines ligand-based descriptors with 3-D spatial information from docked protein-ligand complexes. 3-D spatial information allows our model to learn over the actual binding modality rather than using only structure-based information of the ligand. We show that our model is capable of effectively denoising DEL count data to predict molecule enrichment scores that are better correlated with experimental binding affinity measurements compared to prior works. Moreover, by learning over a collection of docked poses we demonstrate that our model, trained only on DEL data, implicitly learns to perform good docking pose selection without requiring external supervision from expensive-to-source protein crystal structures.

\end{abstract}
\clearpage

\section{Introduction}


One key component of early drug discovery is hit identification, or finding molecules with desired activity levels and properties of interest~\cite{hughes2011principles}. This task has been commonly approached via high-throughput screening (HTS) techniques, which involve testing a library of molecules against a biological target of interest. While traditional screening techniques do not scale well to large chemical spaces, DNA-encoded library (DEL) technologies provide one avenue towards the desired scale.  For instance, while traditional HTS libraries typically contain only $\sim$50k-5M compounds, DELs enable screening of combinatorially large molecular spaces that allow for testing $\sim$1M-5B compounds in a single tube~\cite{satz2022dna, sunkari2021high}. By capturing a broad chemical diversity landscape, DEL technology has opened new opportunities in hit identification~\cite{clark2009design,kleiner2011small,goodnow2017dna,flood2020dna}.



\begin{figure}
    \centering
    \includegraphics[width=.85\linewidth]{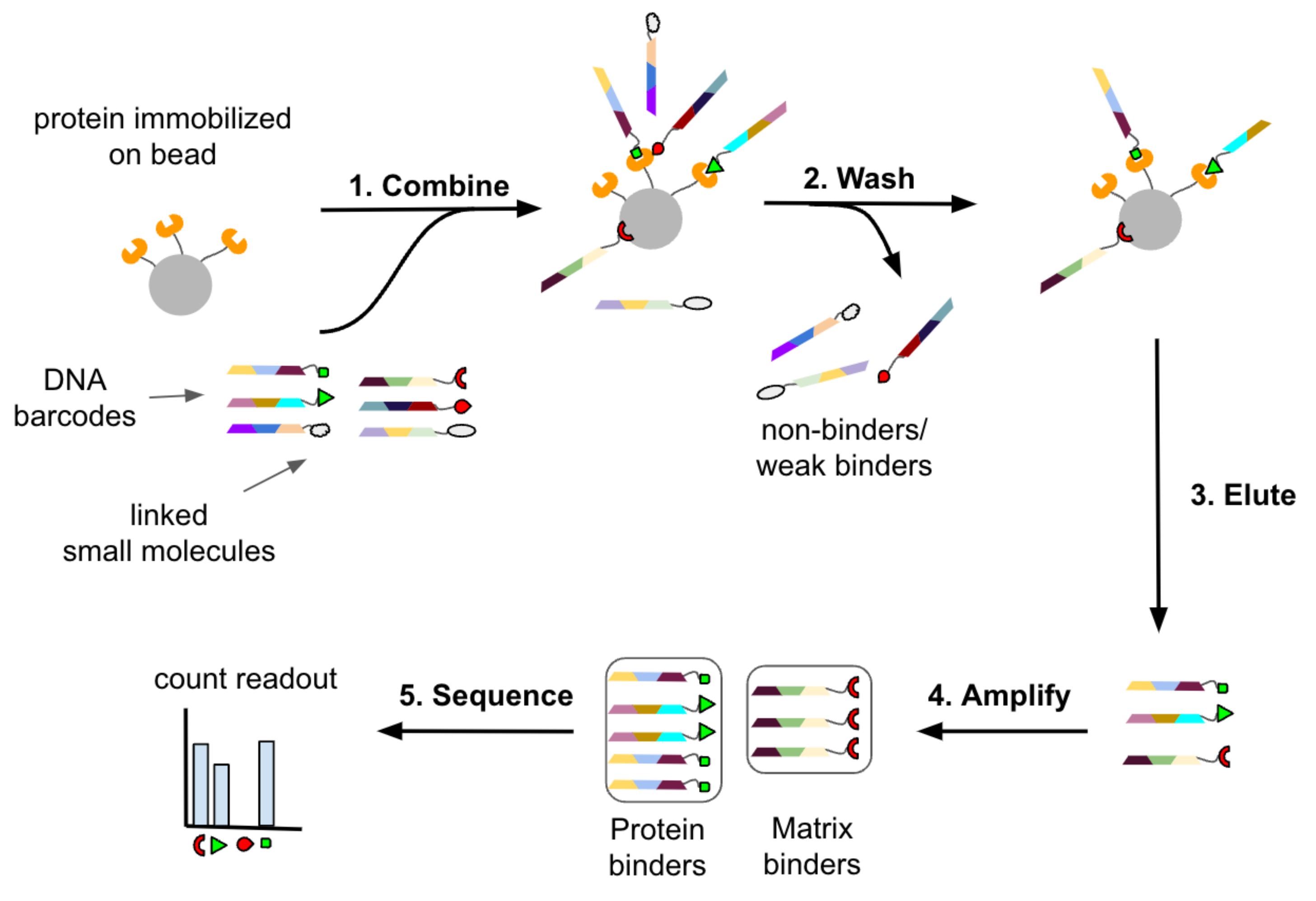}
    \caption{Step-by-step diagram of a DNA-Encoded Library (DEL) panning experiment.  }
    \label{fig:del}
\end{figure}

DELs are constructed by sequentially assembling molecular building blocks, aka synthons, into molecules tagged with unique DNA-barcode identifiers (Fig.~\ref{fig:del}). Once synthesized, the library is tested for affinity against a protein of interest through a series of selection experiments. This process, called panning, typically involves spiking the DEL into a solution of the immobilized protein, and washing the resulting mixture for multiple rounds. This procedure leaves members of the library that remain bound to either the target or matrix, which are subsequently identified using next-generation DNA sequencing. The resulting data after bioinformatics processing consists of sparse reads of the DNA and the corresponding molecules; the relative abundance of the identified molecules is, in theory, a reasonable proxy for their binding affinities. However, DEL data from panning experiments contain various sources of noise such as matrix binding, truncation products, unequal initial loads, replicate and sequencing biases ~\cite{binder2022partial, lim2022machine, zhu2021understanding, komar2020denoising}. Modeling the data without considering these significant sources of noise can overfit to spurious correlations. Consequently, while DELs have proven to be powerful tools in drug discovery, careful consideration of the appropriate techniques to denoise DEL data is required to discover reliable signals of the underlying small molecule binding affinities. 


Several approaches have been developed to tackle DEL modeling from a computational perspective: these existing methods rely on computing an \textbf{enrichment score} for each molecule that is representative of its binding affinity to the protein target. One approach for calculating enrichment scores involves fitting a Poisson distribution to observed on-target and control count data for each molecule and then formulating enrichment scores as ratios of these Poisson parameters~\cite{gerry2019dna}. A deficiency in this enrichment metric is that it neglects to include any structural information of the actual DEL molecules. Since molecular structure represents an important aspect for determining protein-ligand interactions \cite{jones2021improved, hwang2017structure}, other methods incorporate different types of molecular representations into their models. \citet{mccloskey2020machine} use a Graph Neural Network (GNN) to encode molecules, formulating the learning problem as multi-class binary classification over the counts. Recent methods consider hybrid approaches that combine molecular representation learning with probablistic modeling\cite{binder2022partial, lim2022machine, ma2021regression}. Central to these approaches involves modelling observed sequencing counts as originating from latent Poisson or Gamma-Poisson distributions, which are well suited to describe independently distributed count events. This allows the models to incorporate the inherent noise in the data-generating process within the model structure. However, these methods focus on learning only representations of the molecules themselves and omit the rich representation of the 3-D protein-ligand interactions which are critical aspects for describing protein binding.


In order to leverage 3-D spatial data, we turn to molecular docking, which has a long history of being an important tool in structure-based drug discovery \cite{meng2011molecular}. A key component of these docking methods is to define a scoring function that can characterize the binding interactions between a protein and ligand. To generate candidate poses, these methods use the aforementioned scoring function to sample ligand conformations with the goal of ultimately inferring the most likely binding mode. While there have been many advances to docking models and software, in practice, there are still many pitfalls of these methods \cite{chen2015beware}. For instance, the difficulty of the docking problem can be observed through the low empirical correlation between high scoring docked poses and actual binding affinity \cite{gupta2018docking}. Improvements can be realized through carefully calibrated scoring functions, but tuning these scoring functions is frequently problem-specific and involves substantial domain knowledge \cite{li2019overview, pham2008customizing, jain1996scoring}. ML-enabled scoring functions trained on databases of crystal structures have recently demonstrated improved performance in some settings \cite{wallach2015atomnet}, however, these approaches are ultimately limited by scarce and expensive-to-generate crystal structures. Even though docking generates noisy ligand conformations, these docking techniques can provide a useful distribution of likely binding poses which we will  exploit in our new ideas for modeling DEL count data.



Towards the goal of more holistic DEL models, we propose DEL-Dock, a model that directly learns a joint protein-ligand representation by synthesizing multi-modal information within a unified probabilistic framework to learn enrichment scores. Our approach combines molecule-level descriptors with spatial information from docked protein-ligand complexes to explain sequencing counts. We delineate contributions from spurious matrix binding and target protein binding to better separate the signal from noise in the data. Additionally, our model can learn to rank a collection of poses without explicit supervision of pose scores by only training on the count data. When viewed separately, DEL data and docked poses provide noisy signals of binding affinity, but our DEL-Dock model effectively combines these two modalities to better extract the learning signals from the data.


In this work, we engage the commonly studied protein human carbonic anhydrase (CAIX) by training our model on publicly available DEL data generated by \citet{gerry2019dna}. On a held-out evaluation data set of 3041 molecules with experimental affinity measurements extracted from the BindingDB~\cite{liu2007bindingdb} web database, we demonstrate our approach effectively learns enrichment scores that are well correlated with binding affinity. Moreover, we show that our model is capable of extracting insights into the determinants of docked protein-ligand complexes that are most influential for protein binding.   
\section{Methods}





\begin{figure}
    \centering
    \includegraphics[width=\linewidth]{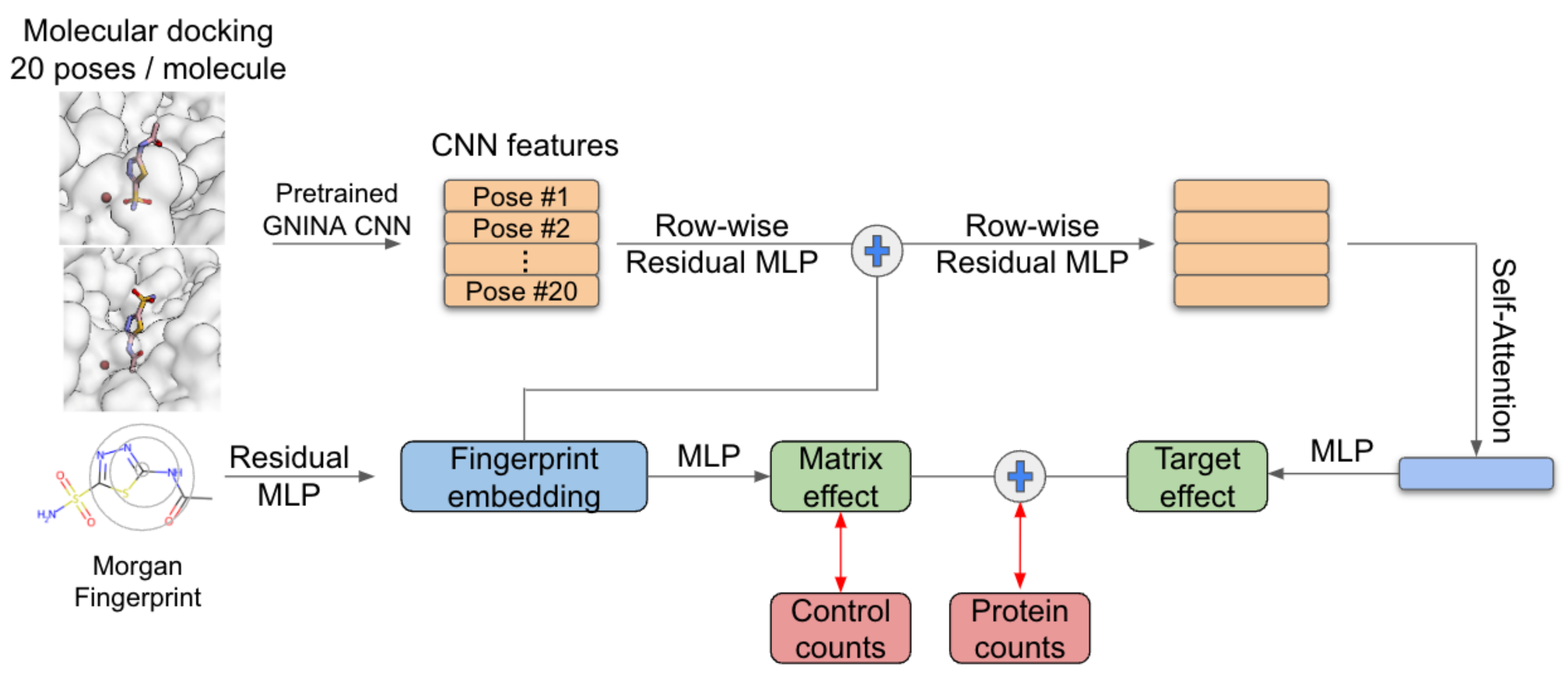}
    \caption{Schematic illustration of our DEL-Dock neural network architechture and data flow.}
    \label{fig:arch}
\end{figure}

Our model combines two different representation modalities, molecule-level descriptors and docked protein-ligand complexes, to capture the latent aspects of protein binding from a probabilistic perspective (Fig.~\ref{fig:arch}). The combinatorial construction of DELs motivates using expressive molecular representations to capture statistical correlation between the building block substructures used in DEL synthesis. We use Morgan fingerprint, calculated with \texttt{RDKit} (version \texttt{2020.09.1})~\cite{greg_landrum_2020_4107869}, as basis for our molecular representations, which is a standard descriptor for representational problems on small molecules~\cite{rogers2010extended} and also provide the added benefit of their simple construction and rapid processing. Morgan fingerprints compute a structural bit hash of the molecule by enumerating $k$-hop substructures about each atom. Since there are many shared structural features across different molecular compounds, these fingerprints constitute a simple representation that has demonstrated remarkable empirical performance throughout cheminformatic domains. We represent docked protein-ligand poses using a pretrained voxel-based CNN model from GNINA, which captures spatial relationships by discretizing space into three-dimentional voxels and leveraging CNNs to learn complex hierarchical representations~\cite{mcnutt2021gnina,ragoza2017protein, sunseri2020libmolgrid}. The CNN models used in this work was originally trained on the PDBBind~\cite{liu2017forging} database, capitalizing on this supervised data source to capture the important features that characterize protein-ligand interactions. 

Let $\mathcal{X}$ denote the set of molecules in our data, where each molecule $x \in \mathcal{X}$ has an associated set of $n$ docked poses $\{p_1, p_2, ..., p_n\} \in \mathcal{P}$ and $c^{\text{matrix}}_i \in C^{\text{matrix}}, c^{\text{target}}_i \in C^\text{target}$ are the $i$th replicates (repeated experiments) of count data from the beads-only control and target protein experiments respectively. Additionally, we can define the following featurization transformations that are used to construct the molecule and pose embeddings: $\Phi : \mathcal{X} \rightarrow [0, 1]^{n_{\phi}}$ is the function that generates a $n_\phi$-bit molecular fingerprint; here, we use a 2048-bit Morgan fingerprint with radius 3. $\Psi : \mathcal{X} \times \mathcal{P} \rightarrow \mathbb{R}^{n_{\psi}}$ is the transformation that outputs an embedding of the molecule and a specific spatial protein-ligand complex, where we use a pre-trained voxel-based CNN to perform this transformation. 

Let $h^{\text{fps}} = \text{MLP}(\Phi(x))$ be the molecule embedding learned by our model, which is computed by applying a multilayer perceptron (MLP) to the fingerprint representation. Individual docked pose embeddings are similarly computed, with one difference being that we also incorporate the fingerprint embedding, $h^{\text{pose}}_i = \text{MLP}([\Psi(x, p_i); h^{\text{fps}}])$ into this representation.

To synthesize the set of poses for each molecule we apply a self-attention layer over the pose embeddings. Following previous work on self-attention and multiple-instance learning (MIL)~\cite{ilse2018attention}, we compute attention weights in \cref{eq:attention}, where $\big(w, W^U, W^V\big)$ are learnable weights, $\sigma$ is the sigmoid activation and $\odot$ is element-wise multiplication. The final output pose embedding that combines information from the individual input poses is then computed as a attention-score weighted embedding vector $h^{\text{pose}} = \frac{1}{n}\sum_i a_i h^{\text{pose}}_i$.

\begin{equation}\label{eq:attention}
    a_i = \frac{ \exp\Big[w \cdot \big(\text{tanh}(W^U h^{\text{pose}}_i) \odot \sigma(W^V h^{\text{pose}}_i)\big) \Big] } {\sum_j \exp\Big[w \cdot  \big(\text{tanh}(W^U h^{\text{pose}}_j) \odot \sigma(W^V h^{\text{pose}}_j)\big) \Big]}
\end{equation}

Equipped with these molecule and pose embeddings, our model learns the contributions of both spurious matrix binding and target protein binding by predicting latent scores that strive to maximize the likelihood of the observed data under the model. We make several distinct modeling choices to mirror our assumptions about the data-generation process that accounts for various sources of experimental noise. 

\textbf{Matrix Binding} is a confounding factor inherent to DEL experiments, since molecules are prone to binding to the multifarious components comprising the immobilized matrix in addition to the intended protein target. For each molecule $x$, we learn a latent matrix binding score $\lambda^{\text{matrix}} = f(h^{\text{fps}})$. Since matrix binding is not a function of the protein-ligand pose representation, we enforce that the matrix binding enrichment remains only a function of the molecule embedding $h^{\text{fps}}$.

\textbf{Target Binding} is learned through $\lambda^{\text{target}} = f(h^{\text{pose}})$, jointly utilizing both molecule and pose representations. This design choice reflects that sequencing counts from the target protein experiment must be a function of both small molecule binding to the protein receptor, represented here as featurizations of the docked protein-ligand complexes, along with promiscuous binding to the immobilized matrix.  


The observed count data for both the control and protein target experiments can be modeled as originating from underlying Poisson distributions, which naturally characterize any discrete count data from independently sampled events. Due to possible sequencing noise we further augment this basic Poisson model as a zero-inflated probability distribution. This design choice is motivated by the chance that sparse zero counts in the data could be explained as an artifact of imperfect sequencing technology, and as a result we directly incorporate this assumption into the structure of our model. We note that previous approaches  have employed Gamma-Poisson distributions to model DEL count data in the past~\cite{ma2021regression}, but we make the assumption here that replicate data for the same molecule is sampled from an identical distribution.

\begin{equation}\label{eq:zip}
\text{P}(C=c | \lambda, \pi) = 
    \begin{cases}
        \pi + (1-\pi)e^{-\lambda} & \text{if } c = 0\\
        (1-\pi)\frac{\lambda^c e^{-\lambda}}{c!} & \text{if } c > 0\\
    \end{cases}
\end{equation}

\begin{align} \label{eq:rate}
    C^{\text{matrix}} &\sim \text{ZIP}(\lambda^{\text{matrix}}, \pi^{\text{matrix}})  &\lambda^{\text{matrix}} &= \exp(\text{MLP}(h^{\text{fps}})) \\
    C^{\text{target}} &\sim \text{ZIP}(\lambda^{\text{matrix}} + \lambda^{\text{target}}, \pi^{\text{target}}) &\lambda^{\text{target}} &= \exp(\text{MLP}(h^{\text{pose}}))
\end{align}

Let $C$ be distributed as a zero-inflated Poisson (ZIP), with its probability density function (PDF) defined in \cref{eq:zip}. Here, $\lambda$ is the rate parameter of the underlying Poisson distribution, and $\pi$ denotes the occurrence of choosing the zero distribution, and is taken to be the empirical average (when $\pi$ = 0, this reduces to the typical Poisson distribution). Empirically, we estimate these zero-count probabilities from the zero-count frequencies in the control ($\pi^{\text{matrix}} \approx 0.0075$) and protein target ($\pi^{\text{target}} \approx 0.55$) experiments.  Since we model the control and protein experiments as originating from separate underlying count distributions, we compute two distinct rate parameters for each ZIP distribution as shown in \cref{eq:rate}. The observed target counts is a function of both matrix binding and binding to the protein target, so the rate parameter for the target distribution is a function of $\lambda^{\text{matrix}}$ and $\lambda^{\text{target}}$. We assume an additive functional form for the latent matrix and target enrichments, which we find works well empirically. We note however that there could be more motivated methods to parameterize the interactions of matrix and target binding. The final loss function is then a typical negative log-likelihood (NLL) loss over the observed counts from both the control and target experiments \cref{eq:loss}. 

\begin{equation} \label{eq:loss}
L = -\sum_{i} \log \big[P(c^{\text{matrix}}_i| \lambda^{\text{matrix}}, \pi^{\text{matrix}}) \big] - \sum_{j}\log[P(c^{\text{target}}_j| \lambda^{\text{target}} + \lambda^{\text{matrix}} , \pi^{\text{target}})]
\end{equation}


\subsection{DEL Data} \label{sec:data}
To train our model we use publicly available DEL data collected by~\citet{gerry2019dna}. This tri-synthon library consists of $\sim$100k molecules  with count data for panning experiments for the human carbonic anhydrase IX (CAIX) protein. In addition to on-target counts, the data includes beads-only no-target controls. Four replicate sets of counts for the protein target experiments are provided, while two replicates of the control experiments are provided in this data set. To account for possible noise in different replicates, we follow previous work and normalize the counts for each target and control replicate by dividing each count by the sum of counts in that replicate experiment and then multiplying by 1e6 to re-calibrate the scale of the counts~\cite{ma2021regression}. This data prepossessing provides the interpretation of each molecule count as a molecular frequency of that molecule within the DEL library. The processed data set is then used to train our models employing a 80/10/10 - train/validation/test split. Complete training details are further provided in the Supporting Information. 


\subsection{Evaluation Data}
We evaluate the performance of our models on benchmarks using an external set of affinity measurements of small molecule curated from the BindingDB~\cite{liu2007bindingdb} web database. We queried binding affinities for the Human Carbonic anhydrase 9 (CAIX) protein target (UniProt: Q16790), and kept only molecules containing the same atom types as those present in the DEL data set (C, O, N, S, H, I). This external evaluation data set is composed of 3041 small molecules with molecular weights ranging from $\sim$25~amu to $\sim$1000~amu and associated experimental inhibitory constant (K$_i$) measurements ranging from $\sim$0.15~M to $\sim$90~pM. We use the median affinity value in the cases where multiple different affinity measurements were reported for the same molecule. We also consider a subset of this dataset which consists of the 521 molecules with molecular weights between 417~amu and 517~amu (Fig.~\ref{fig:mws} in the Supporting Information). These molecular weights correspond to the interquartile range bounding the 10th and 90th percentiles of the molecular weights in training dataset. This restricted subset presents a more challenging test as differentiation cannot rely on only extensive properties such as molecular weight, but must also effectively identify chemical motifs that impact molecular binding within this tightly bound range of molecular weights.  We notice that simple properties such as molecular weight or benzenesulfonamide presence, which is known to be an important binding motif for carbonic anhydrase~\cite{gerry2019dna,li2018versatile,buller2011selection}, achieve better baseline performance on the full evaluation data compared to the restricted subset. These metrics suggest that this subset is more challenging as predictors must learn beyond these simple molecular properties to achieve good performance.

\subsection{Docking}
We perform molecular docking to generate a collection of ligand-bound poses to a target protein of interest for all molecules within in our training and evaluation data sets. Docking is performed using the GNINA docking software~\cite{mcnutt2021gnina,ragoza2017protein, sunseri2020libmolgrid} employing the Vina~\cite{trott2010autodock} scoring function. All molecules are docked against CAIX (PDB:5FL4) with the location of the binding pocket determined by the bound crystal structure ligand (9FK), using the default GNINA settings defining an $8\times8\times8$ \r{A}$^3$ bounding box around this ligand. Initial three-dimensional conformers for all docked molecules were generated with \texttt{RDKit}. For each molecule, we obtain 20 docked poses from GNINA using an exhaustiveness parameter of 50, using the Vina scoring for end-to-end pose generation. This approach can similarly be performed using AutoDock Vina~\cite{trott2010autodock} or Smina~\cite{koes2013lessons} using the Vina scoring function. 

\section{Results}
We demonstrate that our model outperforms previous systems on DEL enrichment prediction by jointly combining topological features from the molecular graph and the spatial 3-D protein-ligand information, and additionally illustrate the capability of our model to better rank ligand poses compared to traditional docking. Our model learns latent binding affinity for each molecule to both the matrix and the target as the denoised signals compared to the observed count data.  In this interpretation we should expect higher enrichment scores predicted by our model to be well-correlated with binding affinity, and therefore provide a useful metric for predicting anticipated protein binding in virtual screening campaigns.

\subsection{DEL-Dock outperforms baselines using only docking or molecule-level descriptors}
To evaluate the performance of our model, especially with respect to out-of-domain protein binding prediction, we first train our model on DEL data screened against the human carbonic anhydrase IX (CAIX) protein target~\cite{gerry2019dna}, and then predict enrichment scores for molecules with externally measured experimental binding affinities to CAIX. We evaluate performance in this setting by measuring spearman rank-correlation coefficients between predicted enrichments and the experimental affinity measurements, which is a metric that is agnostic to the scale of the values. Our model only restricts the enrichment scores to be positive quantities, with no specific distributional constraints, so spearman rank-correlation, which computes a correlation based only on the ordinal ranking of the predicted enrichments, is well suited for our test scenario.


Our method, DEL-Dock, which combines information from docked complexes with molecular descriptors outperforms previous techniques which only utilize one of these two data modalities (see Table~\ref{tab:tab1}). We find that traditional docking scores alone generated from AutoDock Vina~\cite{trott2010autodock} result in the worst overall correlations, commensurate with previous observations that docking scores alone are typically not reliable predictors of binding affinity \cite{gupta2018docking}. Performance based on docked poses alone is however greatly improved when re-scoring the docked poses using pretrained GNINA~\cite{francoeur2020three} CNN models. Another set of baselines we consider are DEL models that rely only on molecular descriptors. First, we consider a simple model that involves training a random forest (RF) on the Morgan fingerprints using the enrichment metrics originally formulated to facilitate analysis of the DEL data set by~\citet{gerry2019dna}. 

We then similarly predict the enrichment scores of molecules in the held-out dataset and the correlation with experimental Ki data. This baseline achieves reasonable performance on both the full and subset evaluation data, especially given the simplicity of the model. For a more sophisticated baseline, we train the Graph Neural Network (GNN) model with the DEL-specific loss function defined by~\citet{lim2022machine} While this approach achieves good performance on the full evaluation data, correlations on the restricted subset are largely unchanged for all docking-based and molecular descriptor-based baselines. Our model which combines docking pose embeddings with molecular fingerprint representations outperforms all other baselines, with the largest improvements of $\sim$2$\times$ better spearman correlations than other approaches realized on the more challenging molecular weight restricted subset. We provide further ablation studies of our model in the Supporting Information.

\begin{table}[]
\begin{tabular}{c|c|c}
\hline
Model                             & Spearman Ki (full) ↓ & Spearman Ki (subset) ↓ \\ \hline
Molecular weight                  & -0.121               & 0.074                  \\ 
Benzenesulfonamide presence                 & -0.199            & -0.063                  \\ \hline
Top Vina~\cite{trott2010autodock} docking score                & -0.068               & 0.119                  \\
Top GNINA~\cite{francoeur2020three} docking score              & -0.279 $\pm$ 0.044   & -0.091 $\pm$ 0.061     \\ \hline
\begin{tabular}[c]{@{}c@{}}RF trained on enrichment \\ scores from (~\citet{gerry2019dna})\end{tabular}& -0.231 $\pm$ 0.007   & -0.091 $\pm$ 0.012     \\ \hline
GNN (~\citet{lim2022machine})                         & -0.298 $\pm$ 0.005   & -0.075 $\pm$ 0.011     \\
\textbf{DEL-Dock (ours)}          & \textbf{-0.328 $\pm$ 0.01}    & \textbf{-0.186 $\pm$ 0.01}
\end{tabular}
\caption{Comparison of spearman rank-correlation coefficients between predicted affinity scores and experimental inhibition constant ($K_i$) measurements curated from BindingDB~\cite{liu2007bindingdb}. Spearman correlations are shown for the complete 3041-molecule data set (full), and a 521-molecule subset of this full data set confined to molecular weights between 417-517~amu. This molecular weight range approximately corresponds to the 10$^{th}$ and 90$^{th}$ interquartile range of the molecular weights spanned by the DEL data set. Error bars are reported as standard deviations over five independently initialized models.}
\label{tab:tab1}
\end{table}


\subsection{DEL-Dock better ranks molecules with known binding chemical motifs}
\begin{figure}
    \centering
    \includegraphics[width=\linewidth]{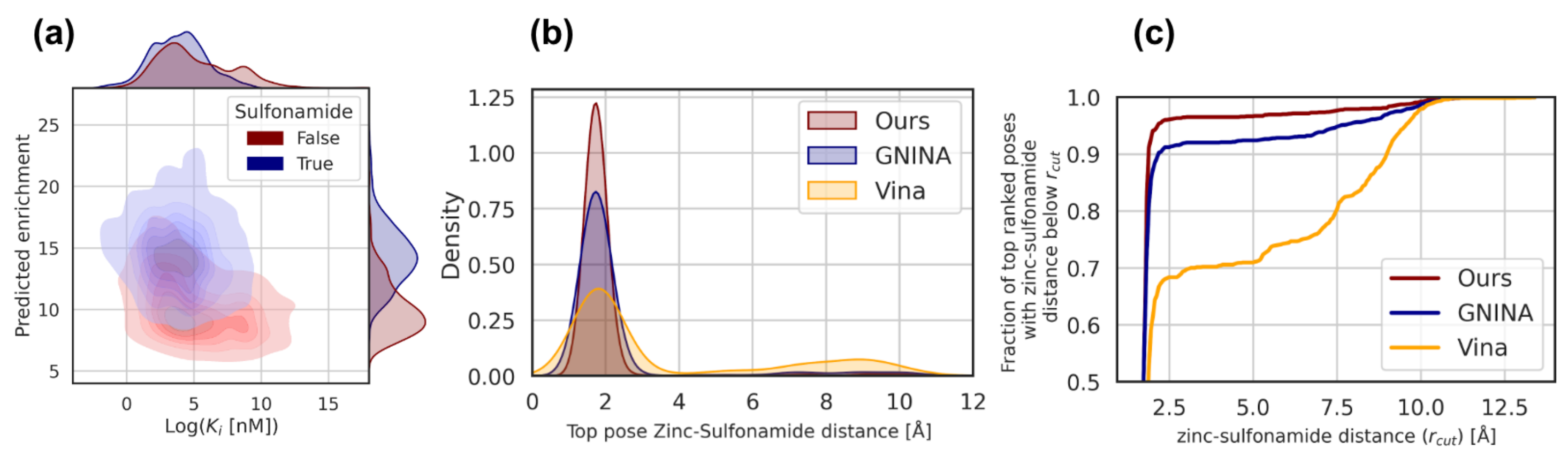}
    \caption{Analysis of DEL-Dock model predictions on our evaluation data set composed of experimental affinity (K$_i$) measurements. \textbf{(a)} Parity plot of our model predicted enrichments and ground truth K$_i$ measurements delineated by benzenesulfonamide presence, with 1581 benzenesulfonamide-containing molecules and 1460 non-benzenesulfonamide. \textbf{(b)} Distribution of zinc-sulfonamide distances for the top-ranked pose of each benzenesulfonamide in our evaluation data set identified by the AutoDock Vina scoring function~\cite{trott2010autodock}, GNINA pose selection~\cite{francoeur2020three}, and our model predicted attention scores. \textbf{(c)} Comparison in the fraction of top ranked poses identified with with zinc-sulfonamide distance below a threshold distance between $\sim$2-12\r{A}. This can be interpreted as the CDF of the distributions in \textbf{(b)} with the appropriate normalization.}
    \label{fig:preds}
\end{figure}

While our approach displays good prediction accuracy with respect to experimental binding measurements, we also find our model provides insights into the structural and chemical factors that influence binding. Compounds containing benzenesulfonamide have been well established in literature as the primarily chemical motif that drives small molecule binding to carbonic anhydrase~\cite{gerry2019dna,li2018versatile,buller2011selection}. Though we do not explicitly incorporate this as a learning signal for our model, we observe that our model is able to learn this association, visibly predicting sulfonamides within our evaluation data set as more highly enriched compared molecules which do not contain benzenesulfonamides (Fig.~\ref{fig:preds}a). Interestingly, we observe a comparatively large fraction of non-benzenesulfonamides identified as good binders with low experimental $K_i$. The elevated population of highly enriched non-benzenesulfonamides in this data set could be an artifact of bias in scientific literature. Our model is ultimately trained on DEL data and therefore is expected to reflect underlying biases and idiosyncrasies of the data generation process. The most notable difference lies in that DEL experiments are only capable of measuring on-DNA binding, while the evaluation data are measurements of off-DNA binding. Nevertheless, the clear delineation of benzenesulfonamides in our predicted enrichments provides good \textit{post-hoc} evidence that our model correctly identifies this important binding motif for this protein target.

An important structural component of benzenesulfonamides binding to carbonic anhydrase is coordination of the sulfonamide group with the zinc ion buried within the active site~\cite{gerry2019dna,li2018versatile,buller2011selection}. In the vast majority of cases, one would then expect docking scoring functions to highly score poses that reflect this anticipated binding mode. As our model performs self-attention over pose embeddings, which are used to learn molecules' enrichment scores, we can interpret the magnitude of the attention probabilities as the importance weight of that particular pose. Shown in Fig.~\ref{fig:preds}b is the distribution of zinc-sulfonamide distances for the top-selected docked pose comparing AutoDock Vina, GNINA, and our method for all 1581 benzenesulfonamides-containing molecules in our evaluation data set. An alternate view of this data is presented as the fraction of top-selected poses with zinc-sulfonamide distances below a distance threshold (Fig.~\ref{fig:preds}c), which can effectively be interpreted as the cumulative distribution function (CDF) of the appropriately normalized associated probability distribution function (PDF) in Fig.~\ref{fig:preds}b. 

The AutoDock Vina scoring function exhibits largest spread of zinc-sulfonamide distances, and as a result identifying a comparatively large fraction of poses as incorrectly coordinated. GNINA pose selection performs significantly better in this setting, identifying a larger fraction of well-coordinated poses with low zinc-sulfonamide distance. We find our method ultimately correctly coordinates the largest proportion of poses when compared to AutoDock Vina or GNINA. We note that our approach for binding pose selection is markedly different than the approach taken by GNINA, which involves a separate pose scoring head trained to identify poses with low-RMSD to ground truth crystal structures. Our attention scores on the other hand are effectively latent variables trained only via the auxiliary task of modeling DEL data. The benefit of our approach is that we can learn to identify good poses in an unsupervised manner, without requiring scarce and expensive crystal structures to serve as the source of supervision for pose selection.

\subsection{DEL-Dock offers interpretability through its attention mechanisms}
\begin{figure}
    \centering
    \includegraphics[width=\linewidth]{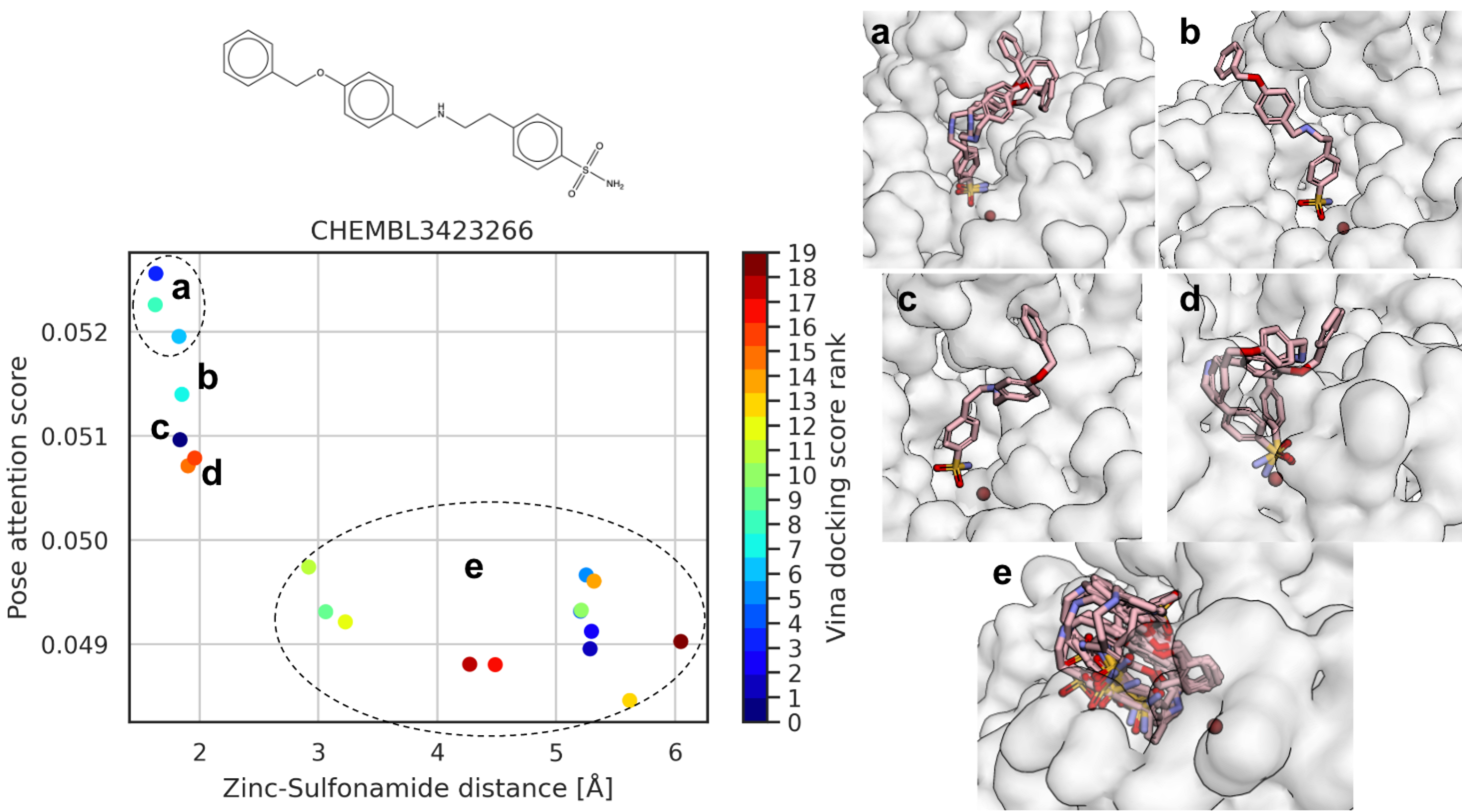}
    \caption{Analysis of pose attention scores for a representative molecule in our evaluation data set. (left) Our model predicted pose attention scores plotted against the zinc-sulfonamide distance of the docked pose and colored according to the ranking determined by the AutoDock Vina scoring function. (right) Different protein-ligand complexes are visualized to show that our model highly ranks the conformers with zinc-sulfonamide coordination (a-d), while the conformers without the correct coordination are ranked lower (e).}
    \label{fig:pose_picture}
\end{figure}

Lastly, we further demonstrate the interpretability of or model by examining the distribution of attention scores learned by our model for a specific molecule (Fig.~\ref{fig:pose_picture}). For this molecule, only 7 out of 20 docked poses correctly coordinate the sulfonamide group with the zinc ion buried in the protein active site. Our model appropriately identifies this binding mode and learns attention scores that more favorably rank these 7 correctly coordinated poses (Fig.~\ref{fig:pose_picture}). The top-three ranked poses (Fig.~\ref{fig:pose_picture}a) by our model have very similar conformations, each exhibiting zinc-sulfonamide coordination, and differing only in the orientation of the terminal benzene ring that is distant from the active site. The other poses that show zinc-sulfonamide coordination (Fig.~\ref{fig:pose_picture}b-d) are also ranked highly by our model, however, these poses exhibit less favorable conformations in several ways. For instance, the conformation in Fig.~\ref{fig:pose_picture}b is more exposed, and less protected by the protein. Finally, our model in general more poorly ranks poses that display incorrect zinc-sulfonamide coordination (Fig.~\ref{fig:pose_picture}e). These conformations typically have the terminal benzene ring inserted into the active site. Also, these poses reveal why zinc-sulfonamide distances alone can be a deceiving metric as some poses are capable of achieving low zinc-sulfonamide distances ($\sim$3~\r{A}) due to the molecule ``curling in'' on itself within the active site. Nevertheless, our model recognizes this spurious binding mode and poorly ranks these poses with comparatively low attention scores, even though AutoDock Vina highly ranks many of these bad poses. Overall, we find the hierarchy of pose rankings by our model to be commensurate with anticipated binding behavior for this protein target.

%
%
%
%
%

\section{Conclusions}


In this work we present an approach for modeling DEL data that combines docking-based and molecular descriptor-based data modalities. Our approach involves predicting two interleaved quantities, enrichment scores, that explain the sequencing counts of the panning experiment measurements for both the on-target protein and the off-target control beads. We evaluate our method by first training on DEL data screened against the human carbonic anhydrase (CAIX)~\cite{gerry2019dna} protein target, and then predicting binding for unseen molecules with external experimental constant of inhibition ($K_i$) affinity measurements curated from the BindingDB~\cite{liu2007bindingdb} web database. For this prediction task we find our approach outperforms previous docking and DEL modeling techniques that only use either docked poses or molecular descriptor information alone. Furthermore, a critical component of our model involves performing self-attention over pose embedding, in order to learn over the set of possible poses. Analyzing these latent attention scores, we find our model effectively identifies good docked poses. Compared to docking pose selection using either AutoDock Vina~\cite{trott2010autodock} or GNINA~\cite{mcnutt2021gnina,ragoza2017protein, sunseri2020libmolgrid, francoeur2020three}, our model more reliably selects poses displaying the appropriate zinc-sulfonamide coordination--which is known to be the predominant binding mode for carbonic anhydrase~\cite{gerry2019dna,li2018versatile,buller2011selection}. Our model is interestingly capable of learning good pose selection in an unsupervised manner, training only on the voluminous DEL data rather than requiring crystal structures to serve as the source of supervision.

There are some limitations to our approach, however. Our approach assumes that we can obtain reasonably accurate protein-ligand docking poses, which is not always true. We can only utilize this model for proteins which have 3-D crystal structures; and even if protein crystal structures are available, they may not always be accurate. Additionally, we are limited by the quality of the docking software, which may fail to capture the actual correct binding modes, or ranking the good binding modes highly. 

Our work focuses on introducing the concept of utilizing docked poses to improve DEL
models, but there are several avenues for future work. While we use Morgan fingerprints as our molecule featurizer, deep learning approaches have the potential to generate more expressive representations, such as Graph Neural Networks (GNNs) that have demonstrated excellent predictive performance on many molecular property prediction tasks \cite{stokes2020deep}. In our application, equivariant GNNs could also be used as an alternative featurizer to CNNs for embedding docked poses, which would provide the added benefit of producing explicitly roto-translationally symmetric representations~\cite{stark2022equibind}. We only select the top ranking poses from docking, but providing a more diverse set of poses could be more useful instead. Future research directions could also leverage our approach to unsupervised pose selection for downstream free energy calculations in larger-scale virtual screening campaigns. Our approach paves the way for multi-modality modeling of DEL data and unsupervised learning of bespoke, DEL-conditioned, scoring functions. 

\section{Acknowledgements}

We would like to thank Nathaniel Stanley and Sam Mun at Insitro for helpful discussions and advice related to Docking software and usage. We would also like to thank Daphne Koller, Robert Hilgraf, Nathaniel Stanley, Fiorella Ruggiu and Yujia Bao at Insitro for providing general feedback and review of our work. Lastly, we would like to thank Patrick Conrad at Insitro for helping us with the public code release.
\section{Data Availability}
We use publicly available data from~\citet{gerry2019dna}, and our code is publicly available at: \url{https://github.com/insitro/insitro-research}.
\newpage
\section{Supporting Information}

\subsection{Training settings}
Featurizations for the docked poses are generated using pre-trained GNINA models provided in \href{https://github.com/RMeli/gnina-torch}{\texttt{gnina-torch}}~\cite{ragoza2017protein,sunseri2020libmolgrid}. The \texttt{dense} variant of the GNINA models composed of densely connected 3D residual CNN blocks introduced in ~\citet{francoeur2020three} are used to generate 224-dimentional embeddings of each docked pose. Morgan fingerprints~\cite{rogers2010extended} for each molecule are calculated using \texttt{RDKit} with a radius of 3 embedded into a 2048 dimensional bit-vector.

All models are trained end-to-end using mini-batch gradient decent with the Adam optimizer~\cite{kingma2014adam} and coefficients for the running averages of $\beta_1=0.95$ and $\beta_2=0.999$. A batch size of 64 is used with an initial learning rate of $1\times10^{-4}$ and a linearly decaying learning rate scheduler where the learning rate is decayed by a factor of $\gamma^{\frac{1}{n_{steps}}}$ every batch. For our learning rate scheduler we use $\gamma=0.1$ and $n_{steps}=1250$, which corresponds to a $10\times$ reduction in the learning rate after $1250$ batches. We also apply gradient clipping, where gradient norms are clipped to a maximum value of 0.1. During training we maintain an exponential moving average over our model parameters which are updated each step with a decay rate of 0.999. This exponential moving average version of the model parameters is then used for evaluation and throughout all inference tasks. Throughout our model we use LeakyReLU activation functions with a negative slope constant of $1 \times 10 ^{-2}$, except for the final activation function applied to the output logits corresponding to the matrix and target enrichment scores where we apply an exponential function as our terminal activation. A hidden dimensionality of 256 is used within MLP layers in our network. The residual MLP layers, which are responsible for processing the Morgan fingerprints along with the CNN features and embeddings (Fig.~2 in the main text), are composed of 2 residually connected MLP layers using dropout with a probability of 0.5. Our model in sum is composed of $\sim$1M parameters and is trained for 8 epochs on a single NVIDIA T4 GPU. A PyTorch implementation of our model that makes use of PyTorch Lightning~\cite{william_falcon_2020_3828935} and Pyro~\cite{bingham2019pyro} is publicly available at: \url{https://github.com/insitro/insitro-research}.

\subsection{Ablations}

We present here a number of ablations on our model, exploring some different architectural components and design choices (Table~\ref{tab:ablate}). First, we see that training models with only fingerprint representations, without incorporating any information from the docked poses, results in a marked decrease in performance. On the other hand models, trained using only CNN representations perform much better and display comparable performance to only GNINA pre-trained models (Table~\ref{tab:tab1}). This represents an intuitive result, as this training setting is effectively equivalent to fine-tuning GNINA using a multi-instance learning over the pose representations. Interestingly, we find that training on the CNN features alone already achieves good binding pose selection based on the latent attention scores, a feature our model is evidently capable of learning in isolation of the fingerprint representations. Also shown as a baseline is the MLP network trained using the bespoke loss function for modelling DEL data presented by \citet{lim2022machine}. This approach represents the lower performing of the two architectures explored by \citet{lim2022machine}, the other being the GNN architecture presented as a baseline in the main text (Table~\ref{tab:tab1}).

\begin{table}[]
\begin{tabular}{c|c|c}
\hline
Model                                                                        & Spearman Ki (full) ↓ & Spearman Ki (subset) ↓          \\ \hline
Only fingerprints                                                            & -0.191 $\pm$ 0.005   & -0.083 $\pm$ 0.019              \\ \hline
Only CNN                                                                     & -0.287 $\pm$ 0.005   & -0.124 $\pm$ 0.006              \\ \hline
MLP from \citet{lim2022machine}                                                                & -0.244 $\pm$ 0.004   & -0.076 $\pm$ 0.017              \\ \hline
\begin{tabular}[c]{@{}c@{}}Without zero-inflated\\ distribution\end{tabular} & -0.26 $\pm$ 0.02     & -0.08 $\pm$ 0.03                \\ \hline
\begin{tabular}[c]{@{}c@{}}End-to-end voxels\\ with frozen CNN\end{tabular}  & -0.278 $\pm$ 0.022   & -0.16 $\pm$ 0.03
\end{tabular}
\caption{Model ablations and other baselines. Error bars are calculated as standard deviations over five independently initialized models.}
\label{tab:ablate}
\end{table}

Interestingly using a zero-inflated loss appears to be critical to our performance, resulting in $\sim$25\% increase in spearman rank-correlation on the full evaluation set and greater than a 2$\times$ increase on the subset. We suspect this performance jump could be related to the disparity in zero-counts between the control and on-target experiment: the control experiments have a zero-count frequency of $\sim$0.75\% while the protein target experiments have a zero-count frequency of $\sim$55\%. Using a zero-inflated distribution could provide our model more flexibility to explain zero-counts as an artifact of the data generation process, rather than an outcome of poor protein binding.

Instead of using pre-computed CNN features from GNINA we also explored training our model directly from the voxel representations using frozen CNN featurizers. The benefit of this approach is the ability to use data augmentation via radom rotations and translations to implicitly enforce that the learned CNN embeddings remain roto-translationally equivariant. While we notice performance on the evaluation subset is comparable with our trained on pre-computed CNN features (Table ~\ref{tab:tab1}), the performance on the full data set is slightly reduced. We suspect this result could be due to the computational challenges of using voxelized representations. In particular, when training over many docked poses (in our case 20 poses per molecule) our batch size is effectively 20$\times$ larger -- which presents a significant memory bottleneck as the voxel representation requires storing a 48$\times$48$\times$48$\times$28 molecular grids (three dimensions discretizing space, and one for different atom types). Furthermore, our pre-computed features are already being generated with pre-trained CNN featurizers that have been trained using data augmentation, albeit on PDBBind for the separate task of affinity and pose prediction. Nevertheless, we certainly expect improved performances could still be achieved for these full differentiable training approaches given the appropriate compute resources and further tuning.

Lastly, presented in Table~\ref{tab:poses_ablate} is a comparison of spearman rank-correlation performances training on variable numbers of poses. For each model the top-$k$ poses generated via docking are used for training. Performance tends to generally improve with increasing number of poses used for training, with the largest difference in improvements realized on the molecular weight restricted subset. Beyond $\sim$10 poses appears to result in diminishing returns, in comparison to the jumpy in improvements seen from 2 $\rightarrow$ 10 poses. 

\begin{table}[]
\begin{tabular}{c|c|c}
\hline
Number of training poses                                                                        & Spearman Ki (full) ↓ & Spearman Ki (subset) ↓          \\ \hline
2 poses                                                            & -0.278 $\pm$ 0.011   & -0.112 $\pm$ 0.023              \\ \hline
5 poses                                                                     & -0.304 $\pm$ 0.01   & -0.15 $\pm$ 0.02              \\ \hline
10 poses                                                                & -0.318 $\pm$ 0.007   & -0.175 $\pm$ 0.02              \\ \hline
15 poses & -0.324 $\pm$ 0.008     & -0.182 $\pm$ 0.014                \\ \hline
20 poses  & -0.328 $\pm$ 0.009   & -0.186 $\pm$ 0.013
\end{tabular}
\caption{Model ablations training on different numbers of docked poses. Error bars are calculated as standard deviations over five independently initialized models.}
\label{tab:poses_ablate}
\end{table}

\subsection{Supplementary Figures}

\begin{figure}
    \centering
    \includegraphics[width=.7\linewidth]{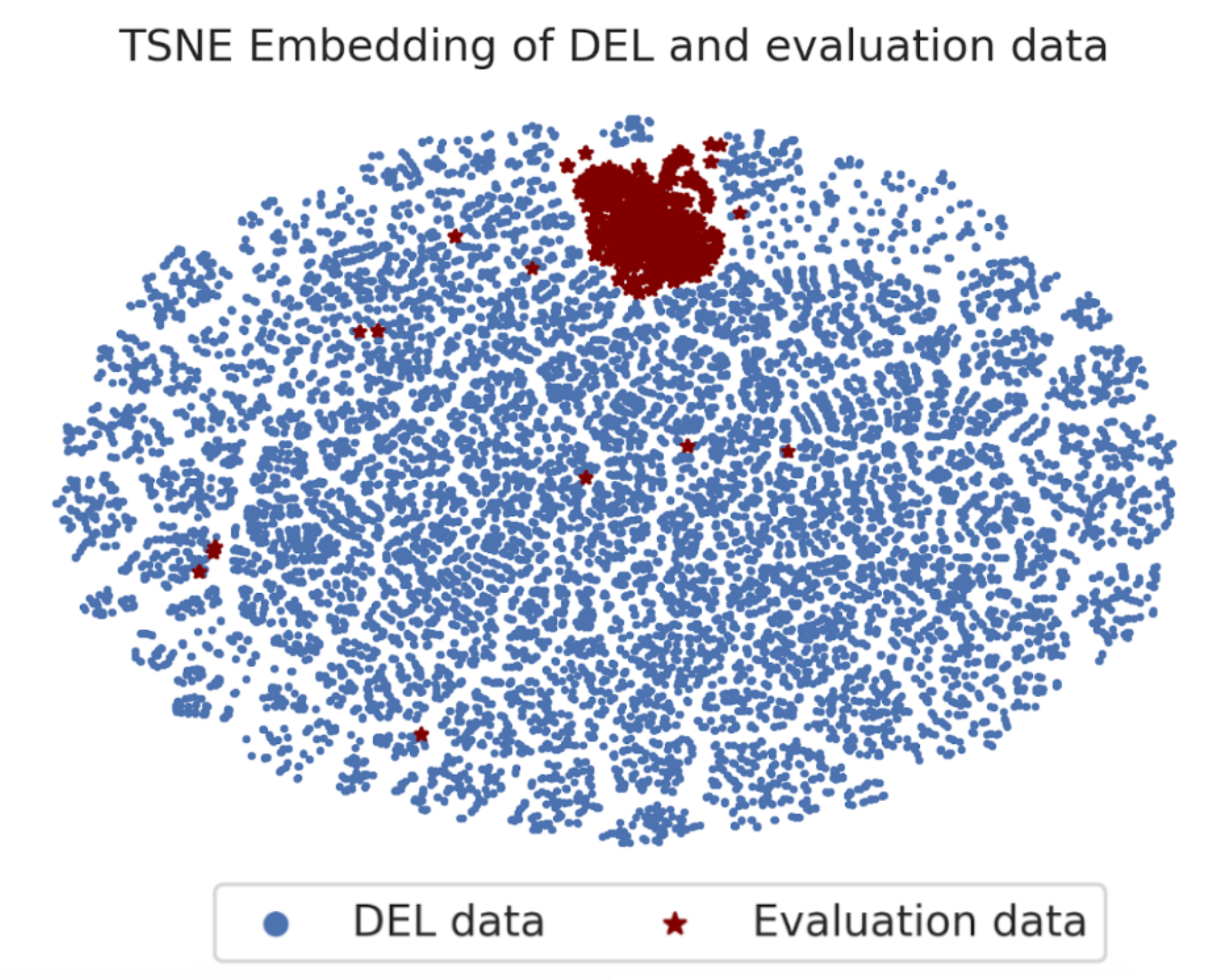}
    \caption{TSNE embedding our of DEL and evaluation data. Molecular representations for this TSNE embedding are generated as a concatenation of three fingerprint representations: Morgan fingerprints, MACCS fingerprints, and atom pair fingerprints.}
    \label{fig:tsne}
\end{figure}

\begin{figure}
    \centering
    \includegraphics[width=.9\linewidth]{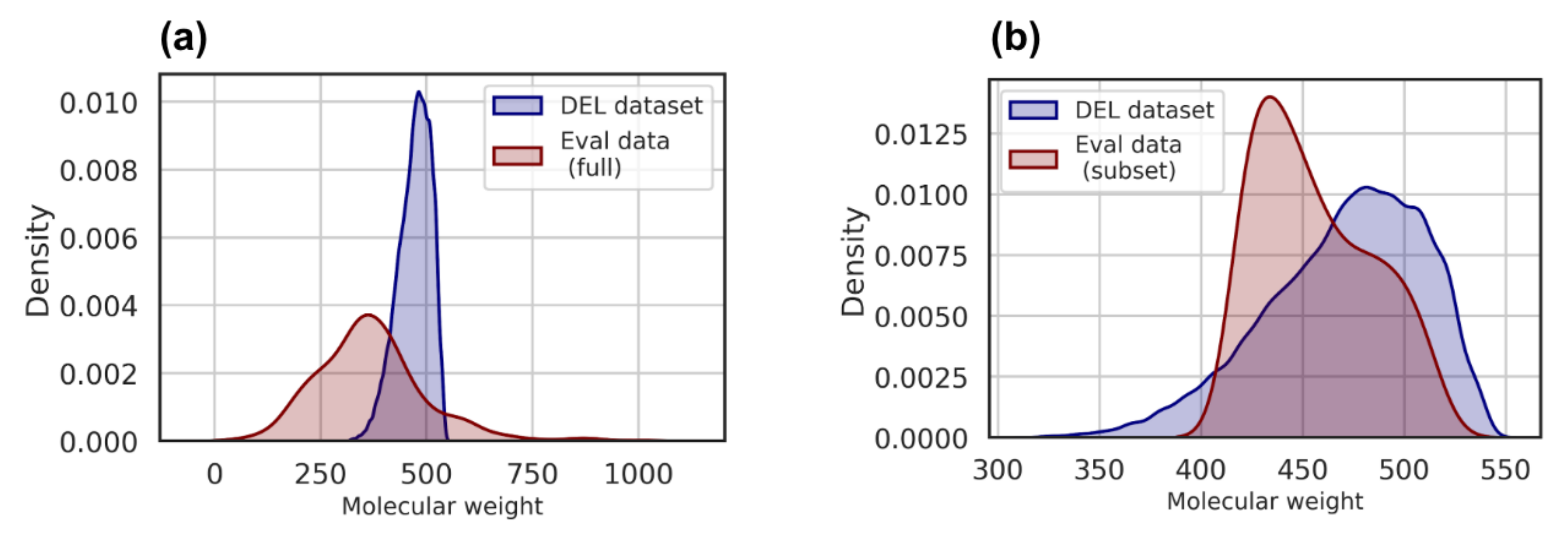}
    \caption{Comparison of the distribution of molecular weights between the DEL data set and \textbf{(a)} the full evaluation data set and \textbf{(b)} the 417-517 amu subset of the evaluation data set. Distributions are generated as a Kernel Density Estimate (KDE) plot as implemented in \texttt{seaborn}~\cite{Waskom2021}.}
    \label{fig:mws}
\end{figure}

Shown in Fig.~\ref{fig:tsne} is a TSNE embedding of the DEL data set alongside our evaluation data. This TSNE embedding is generated by representing each molecule with a concatenation of three fingerprint representations: a 2048-dimensional Morgan fingerprint with a radius of 3, 167-dimensional MACCS (Molecular ACCess System) fingerprint, and finally a 2048-dimensional atom pair fingerprint. All fingerprints are calculated using \texttt{RDKit}. Sci-kit learn is then used to generate the TSNE embedding using a tanimoto similarity metric with a perplexity of 30 trained on the combined DEL and evaluation data. We notice the evaluation data is largely isolated from the DEL data in this TSNE embedding, serving as an indication that our evaluation data is markedly different, or out of domain, than the DEL data used in training our models.      

\begin{figure}
    \centering
    \includegraphics[width=\linewidth]{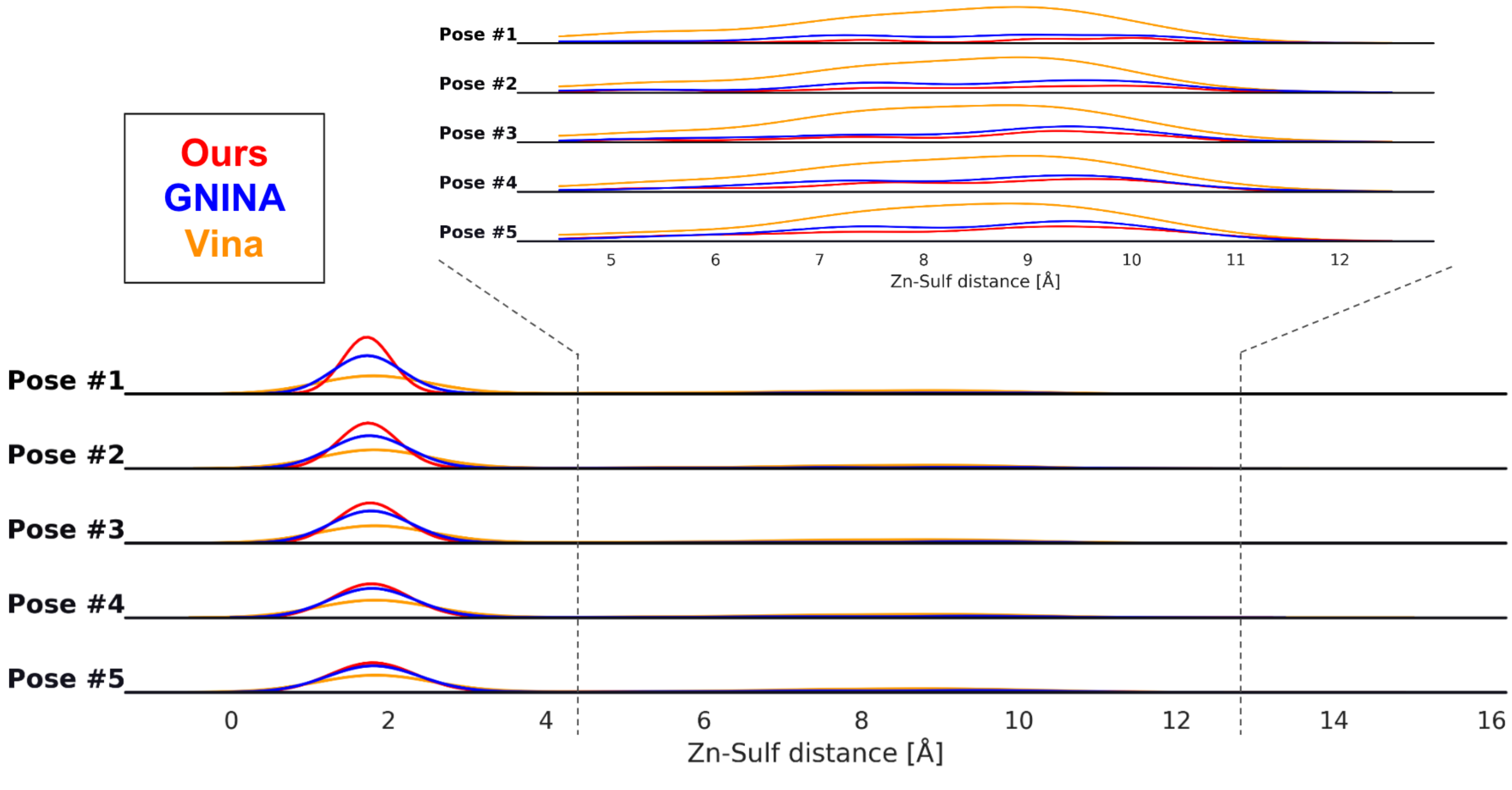}
    \caption{Distribution of Zinc-Sulfonamide distances throughout the top-five ranked poses by our DEL-DOCK model attention scores, GNINA pose selection score, and AutoDock Vina scoring function.}
    \label{fig:top_poses}
\end{figure}

Shown in Fig.~\ref{fig:top_poses} are the distributions of zinc-sulfonamide distances throughout the top-five ranking poses as identified by our DEL-DOCK model attention scores, GNINA pose selection, and the AutoDock Vina scoring function. We notice that the highly ranked poses by our model attribute more density in the closely separated regime under $\sim$4\r{A} than GNINA or Vina, and as a direct consequence of this we see fewer poses selected by our model showing large separations between $\sim$4\r{A} - $\sim$13\r{A}. 

%
%

\bibliography{references}

\end{document}